\documentstyle[floats,epsf,epsfig,prbbib,aps]{revtex}

\begin{document}

\twocolumn[\hsize\textwidth\columnwidth\hsize\csname
@twocolumnfalse\endcsname

\title{\vspace*{-1cm}\hfill
{\tt Phys.~Rev.~B in print}
       \vspace{1cm}\\
The influence of surface stress on the equilibrium shape of strained 
quantum dots}

\author{N.~Moll \cite{Author} and M.~Scheffler} 
\address{ Fritz-Haber-Institut der Max-Planck-Gesellschaft, Faradayweg 4-6,
D-14195 Berlin-Dahlem, Germany}
\author{E.~Pehlke} 
\address{Physik-Department T30, Technische Universit{\"a}t M{\"u}nchen,
D-85747 Garching, Germany}
\date{\today}
\maketitle

\begin{abstract}
The equilibrium shapes of InAs quantum dots ({\em i.e.}, dislocation-free,
strained islands with sizes $\ge 10,000$ atoms) grown on a GaAs (001)
substrate are studied using a hybrid approach which combines density
functional theory (DFT) calculations of microscopic parameters, surface
energies, and surface stresses with elasticity theory for the long-range
strain fields and strain relaxations. In particular we report DFT calculations
of the surface stresses and analyze the influence of the strain on the surface
energies of the various facets of the quantum dot. The surface stresses have
been neglected in previous studies. Furthermore, the influence of edge
energies on the island shapes is briefly discussed. From the knowledge of the
equilibrium shape of these islands, we address the question whether
experimentally observed quantum dots correspond to thermal equilibrium
structures or if they are a result of the growth kinetics.
\end{abstract}

\vskip2pc]

\thispagestyle{empty}

\section{Introduction}

In recent years, the study of growth conditions and electronic properties of
quantum dots has attracted significant attention in basic science and
technology.
\cite{guha:90,leonard:94,moison:94,ruvimov:95,grundmann:95,georgsson:95,ledentsov:96a}
Quantum dots are small three-dimensional islands of a low-band-gap
semiconductor (e.g. In$_x$Ga$_{1-x}$As) which are enclosed in a wide-band-gap
semiconductor matrix (e.g. GaAs). Provided the bands of these islands and of
the host are appropriately aligned, the valence and conduction bands produce a
confinement potential for the holes in the valence band and the electrons in
the conduction band. If these islands are small enough, they will behave like
big artificial atoms with discrete energy levels. Thus, the recombination
spectrum of a single quantum dot consists of a single sharp line with a
practically not measurable temperature broadening.  Quantum dots may be used
for new types of devices, as for example a single-electron transistor or
cellular automata. Further examples are semiconductor lasers, where the wave
length of the emitted light is determined by alloy composition and the size
and shape of the dots. Indeed, such lasers have been build in the
laboratory. \cite{grundmann:95,georgsson:95} The required size of the quantum
dot is dictated by the condition that the energy separation of the quantized
electronic levels of the dots should be about $0.1 - 0.2$ eV, so that they are
not populated at room temperature. And the dot has to be sufficiently large
that at least one bound level exists; for too small islands this is not the
case. \cite{ledentsov:96} For GaInAs these conditions imply that the width of
the island is between 50 and 200 {\AA} (or $20 - 80$ atoms), which means that
we are dealing with objects built of $1,000 - 60,000$ atoms. For lasers it is
necessary to have many dots ($ \sim 10^{11} {\rm cm}^{-2}$) and these should
all have nearly the same size and shape, so that all dots emit light at
practically the same wave length; in a quantum dot laser the width of the line
is determined by the size and shape fluctuations of the ensemble of quantum
dots which implies that the size and shape uniformity of quantum dots is
critical to these applications.

Already in 1990 it had been observed that dislocation-free, strained islands
form by itself when InAs is deposited on GaAs, \cite{guha:90} and since 1994
several groups \cite{leonard:94,moison:94,ruvimov:95} have shown that a range
of growth parameters exists at which quantum dots assemble themselves with the
desired and tunable size and a rather narrow size distribution. The mechanism
giving rise to this self-assembly of the dots is still not understood.

Often the formation of quantum dots is explained in terms of a
thermal-equilibrium picture where the system assumes the state of lowest free
energy: Islands form, instead of a strained, epitaxial film, because the gain
of elastic relaxation energy (possible in an island) over-compensates the cost
due to the increased surface energy (a three-dimensional island has a larger
surface than a two-dimensional film).  Typically such island do not form
immediately on the substrate but on top of a wetting layer (see for example
Ref. \onlinecite{ledentsov:96}).  For InAs quantum dots on GaAs (001) this
wetting layer has a thickness of about 1.5 monolayers. When this thermal
equilibrium picture applies the growth mode giving rise to islands is called
the Stranski-Krastanov growth mode. \cite{stranski:37}

Other authors have emphasized the role of kinetic effects.
\cite{jesson:96,chen:96,dobbs:97} Dobbs {\em et al.}~\cite{dobbs:97} studied
the formation of islands using self-consistent rate equations. Their rate
theory is designed to predict reliably average quantities and they found that
their island density in dependence of the coverage is in good agreement with
experiment. But they could not calculate the island size distribution
explicitly.

Because in several experimental as well as theoretical papers
\cite{ledentsov:96a,shchukin:95,shchukin:96} it is assumed that
thermal-equilibrium theory is applicable to describe and understand the
formation of quantum dots, we performed calculations of the equilibrium
structure to find out whether or not this agrees with results of growth
experiments. The calculations are done for dislocation-free InAs islands
epitaxially grown on GaAs (001). Knowing the equilibrium shapes
is indeed important, because under certain conditions thermal equilibrium will
be reached.  And in general, if the quantum-dot shape observed in experiments
deviates from the equilibrium shape, one has to conclude that equilibrium
thermodynamics is not adequate to describe the island formation and the
assumed shape and size distribution.  We note in passing that the growth of
experimental quantum dots is likely affected also by (unwanted) alloying of
the dot and the matrix, \cite{rosenauer:97} by entropy effects, and for high
concentrations of quantum dots also the island-island interaction has to be
considered. These effects are neglected in the present calculations.

\begin{figure}[tb]
    \epsfxsize=\linewidth
    \epsfbox{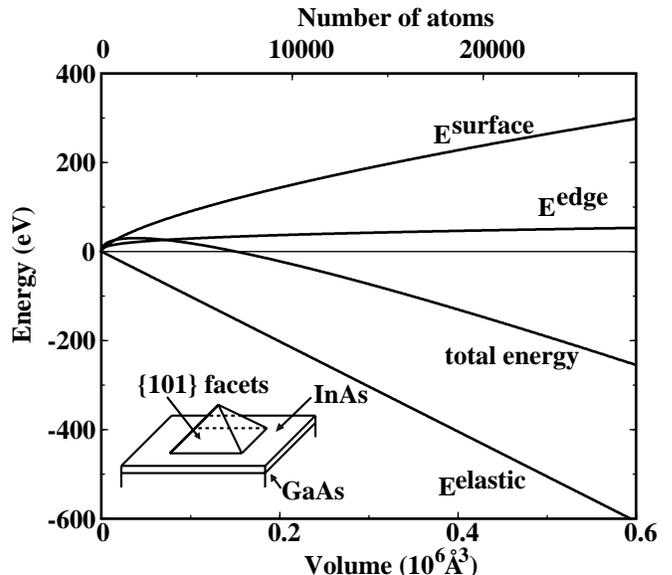}
\caption{Energy contributions of a single, pyramidal shaped, dislocation free
  island compared to the energies of an epitaxial (strained) film. The elastic
  energy relief $E^{\rm elastic}$ due to the partial strain relaxation, the
  surface-energy contribution $E^{\rm surface}$ due to the increased surface
  area, and the contribution due to the various edges $E^{\rm edge}$ are
  shown. The total energy displays the sum of all these contributions. The
  applied approach (see text) is valid for island sizes bigger than 1,000
  atoms. The elastic constants used for the strain relaxation are given in
  Tab.\ \ref{emodul}. The surface energies stresses are those of Tab.\
  \ref{energies}. For the edge energy we used $\gamma^{\rm edge}=50$
  meV/{\AA}.}
  \label{scaling}
\end{figure}

The InAs quantum dots of interest contain more than 1,000 atoms. We consider
it unwise to evaluate the total energy of such a system by a {\em direct}
density functional calculation. Therefore a hybrid method was developed
that provides the same result as a direct approach, but at much less
computational costs. Even more importantly, this method exhibits with greater
clarity the underlying physical mechanisms.  In brief, the approach is
summarized as follows. The total energy of a large, isolated quantum dot is
given by
\begin{equation}
E^{\rm q-dot} =  E^{\rm elastic} + E^{\rm surface} + E^{\rm edge} \quad.
\label{q-punkt}
\end{equation}
The leading terms are the elastic relaxation energy and the sum over the
surface energies of the surface facets. Both quantities depend sensitively on
the quantum-dot shape. The surface reconstructions, the surface energies and
their strain dependence are calculated by DFT and analyzed as a function of
the atomic chemical potential.  If the size of an island is bigger than about
1,000 atoms it turns out that the strain fields and elastic energies are well
described by elasticity theory because they follow the scaling
laws. \cite{priester:95} We therefore evaluate the long-range strain
relaxation in the quantum dot and in the underlying substrate by elasticity
theory applying a finite-element approach.  The approach permits the
systematic investigation of almost any island shape. Fig.\ \ref{scaling}
displays the importance of the different energy contributions and their
scaling with island size. These results refer to an isolated, pyramidal
shaped quantum dot, but essentially the same behavior is found for other
island shapes.  The elastic energy of a relaxed quantum dot, compared to the
energy of a two-dimensional epitaxial film scales linearly with the volume (or
the number of atoms) of the quantum dot.
This analytical scaling holds true for the elastic energy as long as the
islands contain more than 1,000 atoms. The single facets of the island have to
be larger than $\sim 16$ atoms, so that the reconstructions on the facets are
not suppressed and therefore the surface energy scales then with the area.
This implies it scales with the volume to the power 2/3. Surface energy is a
cost, and therefore the contribution is positive. Also shown is the
contribution from the edges which also is a cost, though a rather small
one. Obviously, the sum over the energies of the edges scales with the volume
to the power 1/3. So that the edge energies provide the scaling relation they
should be larger that $\sim 4$ atoms. The main uncertainty in using the above
approach is that there are no edge energies known until now and they only can
be estimated. Furthermore, for small islands the atomic structure on the side
facets might not reconstruct.

This method of calculating the total energy of dislocation-free, strained,
relaxed islands (as described by Eq.\ \ref{q-punkt}) was used before
\cite{pehlke:96,pehlke:97} but so far some approximations were implied which
will now be dropped.  The approximations were: $i)$ the elastic properties of
the InAs islands were taken identical to those of the GaAs substrate, $ii)$
the influence of the surface strain on the surface energies of the various
facets was neglected, $iii)$ the influence of the edge energies was not
discussed. The improved treatment reported in this paper required elaborate
calculations (this applies in particular to the surface stresses), but our
results fully confirm the earlier conclusions. The obtained quantitative
differences to the previous work are small.

In the following Section we discuss some results of the finite element
calculations.  Then, Section III presents the calculations of surface energies
and surfaces stresses for the low-index surfaces: (001), (110), (111), and
(\=1\=1\=1). For the (110) surface we also give the results for the first
derivative of the surfaces stress with respect to strain.  Combining the
results for elastic and surface energies we obtain the total energy of the
islands, and from the total energies of all different island shapes we derive
the energetically favorable island shape for a given volume. This analysis is
done in Section IV, where we also discuss the influence of the edge energies.

\section{Long-Range Strain Relaxation in the Island and the Substrate}

We have calculated the elastic energy within the continuum theory. The
experimental elastic moduli (see Tab.\ \ref{emodul}) are used to describe 
the elastic properties of both the substrate and the island.   
\begin{table}[b]
  \begin{center}
    \begin{tabular}{lrrr}
      material & $c_{11}$ [GPa] & $c_{12}$ [GPa] & $c_{44}$ [GPa] \\ \tableline
      GaAs     &   119          & 53.8           & 59.4 \\
      InAs     &  83.3          & 45.3           & 39.6 \\
    \end{tabular}
  \end{center}
  \caption{The experimental elastic moduli $c_{11}$, $c_{12}$ and $c_{44}$ of
    GaAs and InAs. \protect\cite{landolt:82}}
  \label{emodul}
\end{table}
A finite element approach is applied to solve the elasticity problem. Both the
island and a sufficiently thick slab (240 {\AA} for a quantum dot volume of
$2.88 \times 10^5$ {\AA}$^3$) representing the substrate are divided into
small irregularly shaped tetrahedra. The displacement field is tabulated on
the vertices of this partitioning. Within each tetrahedron the linear
interpolation of the displacement field is uniquely determined by the values
at the four corners of the tetrahedron. The total elastic energy is calculated
by summing the elastic energy density within each tetrahedron, which is a
function of local strain, times the volume of the unstrained tetrahedron over
all tetrahedra.  This expression is iteratively minimized with respect to the
displacement field. This procedure is repeated for finer and finer
partitioning of the volume and the results are finally extrapolated to
fineness equal to zero.

\begin{figure}[tb]
  \begin{center}
    \epsfig{figure=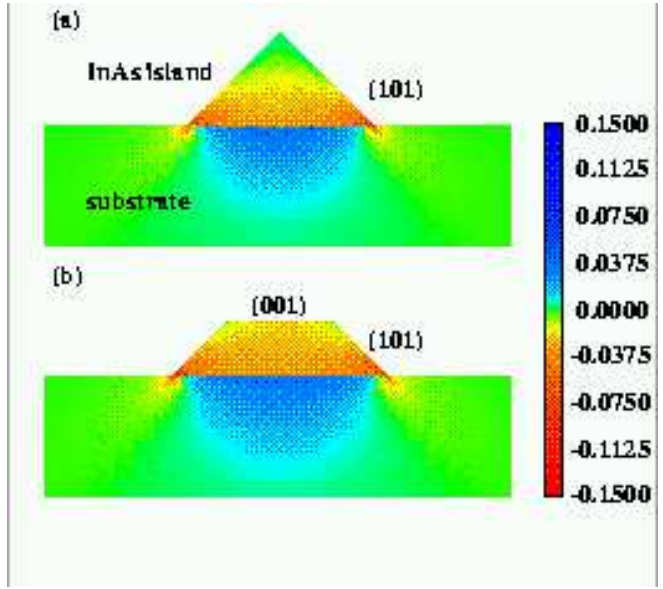, width=\linewidth}
  \end{center}
  \caption{The trace of the strain tensor for (a) a pyramidal and (b) a
   truncated island on the (010) cross sections through the islands. Note that
   the actual calculation has been carried out for a much thicker slab.}
  \label{strain}
\end{figure}

To obtain the elastic energy of truncated islands we use a simple analytic
approximation to avoid repeating the full finite-element calculation. This
analytic expression is based on the scaling law for the elastic energy and on
the fact that the tops of the pyramidal islands are almost completely
relaxed. The islands relaxes about 50 \%, if the substrate is kept fixed.
Additional 15 \% are gained if both island and substrate are relaxed. In Fig.\
\ref{strain} the strain field of a pyramidal and a truncated island are
compared. For both shapes the trace of the strain tensor is shown on a (010)
cross section plane.
The top of the pyramidal island is almost fully relaxed and therefore the
elastic energy is almost completely stored in the base. From this
observation and the scaling law of elastic energy with volume, one easily
can derive the analytic approximation for the elastic relaxation energies of
truncated pyramids which was proven to be sufficiently
accurate. \cite{pehlke:97} 

\section{Surface Energies, Surface Stresses, and Their First
Derivatives}

The InAs surface energies and surfaces stresses are calculated
\cite{bockstedte:97} using density-functional theory and the local-density
approximation for the exchange-correlation energy functional. \cite{perdew:81}
We use {\em ab initio}, norm-conserving, fully separable pseudopotentials.
\cite{hamann:89,fuchs:98,kleinman:82} The wave functions are expanded into
plane waves with an energy cutoff of 10 Ry. The {\bf k}-summation is done by
using a uniform Monkhorst-Pack mesh \cite{monkhorst:76} with a density
equivalent to 64 {\bf k}-points in the whole $(1 \times 1)$ surface Brillouin
zone of the (100) surface. To obtain the absolute surface energies for (111)
and (\=1\=1\=1) orientations we employ the energy-density formalism introduced
by Chetty and Martin. \cite{chetty:92a} Corresponding calculations were done
before for GaAs and are described in Ref.\ \onlinecite{moll:96}. As the InAs
surface reconstructions are similar to those of GaAs, \cite{fawcett:93} we
choose the same candidates for the low-energy surface structures. Indeed, we
find the same behavior, except that As-rich reconstructions like the As
terminated (110) surface are energetically unfavorable and not
thermodynamically stable.

\begin{figure}[tb]
  \epsfxsize=\linewidth
  \epsfbox{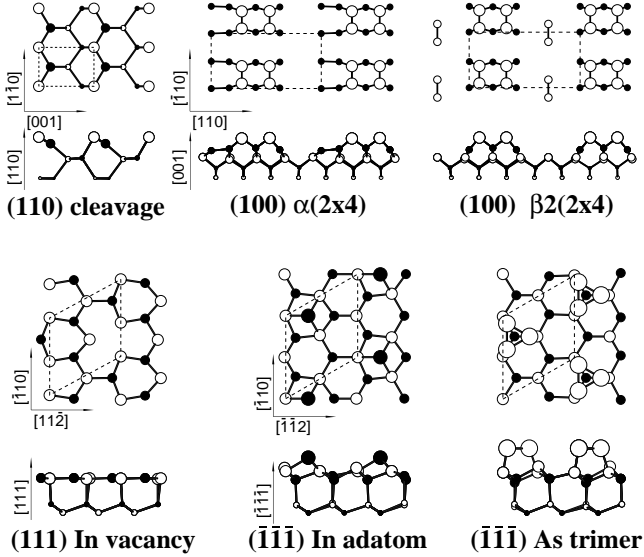}
  \caption{Atomic structure models for the different InAs surfaces, top and
    side views. Filled and open circles denote In and As atoms, respectively.}
  \label{geometry}
\end{figure}

\begin{figure}[tb]
  \epsfxsize=\linewidth
  \epsfbox{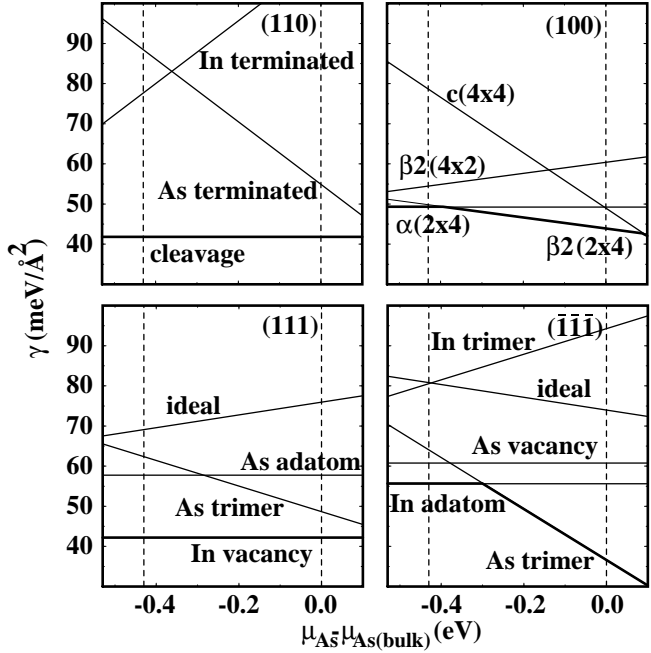}
  \caption{InAs surface energies of the (110), (100), (111), and (\=1\=1\=1)
    surface orientation as a function of the As chemical potential. The thick
    lines highlight the calculated surface energies of the 
    reconstructions of lowest total energy. }
  \label{energies}
\end{figure}

The relaxed atomic surface geometries of the equilibrium structures
are displayed in 
Fig.\ \ref{geometry}.
In Fig.\ \ref{energies} the surface energies are shown as function of the As
chemical potential. The left and right vertical dashed lines denote a In and
As-rich environment, respectively.
For the (100) orientation the $\alpha(2 \times 4)$ and $\beta 2(2 \times 4)$
reconstructions have the lowest surface energy. Both the (110) and (111)
surface energies are independent of the As chemical potential.  The relaxed
$(1 \times 1)$ cleavage surface is the stable reconstruction for the (110)
orientation and the In vacancy structure for the (111) orientation. On the
(\=1\=1\=1) surface an As trimer reconstruction forms in As-rich
environment. In In rich environment the In adatom structure is energetically
preferred.  We note that a $(\sqrt{19} \times \sqrt{19})$ structure had been
observed in the case of GaAs by scanning tunneling
microscopy. \cite{biegelsen:90} However, we have not yet carried out
calculations for this reconstruction, which would be rather expensive due to
the large unit cell.

Because epitaxial growth is often performed under As-rich conditions, we
present in Tab.\ \ref{tab} the surface energies for $\mu_{\rm As} = \mu_{\rm
As(bulk)}$, {\em i.e.}, $\mu_{\rm As}$ is taken at the value of the right
dashed line of Fig.\ \ref{energies}.
\begin{table}[tb]
  \caption{Surface energies $\gamma$ and surface stresses $\sigma_x$, 
    $\sigma_y$ for InAs surface reconstructions in equilibrium with bulk As.}
  \begin{tabular}{clcccc}
    \multicolumn{2}{l}{surface} & $\gamma$ & $\sigma_x$ & $\sigma_y$\\
    &        & (meV/{\AA}$^2$) & (meV/{\AA}$^2$) & (meV/{\AA}$^2$)\\
    \tableline
    (110) & cleavage             & 41 & 26 & 54\\
    (100) & $\beta2(2\times 4)$  & 44 & -- & --\\
    (111) & In vacancy           & 42 & 48 & 48\\
    (\=1\=1\=1) & As trimer      & 36 & 92 & 92\\
  \end{tabular}
  \label{tab}
\end{table}
Furthermore, we have calculated the surface stress for the reconstructions
stable under As-rich conditions, and we calculated the strain derivatives of
the stress.  These results are required in order to obtain the corrections of
the surface energy for strained systems. The surface energy of a strained
surface defined with respect to the area of the undeformed surface is given by 
\begin{equation}
\gamma^{\rm strained} = \gamma
+ \sum_{ij} \sigma_{ij} \epsilon_{ij}
+ \frac{1}{2} \sum_{ijkl} \epsilon_{ij} S_{ijkl}
  \epsilon_{kl} + \ldots \quad,
\end{equation}
where $\gamma$ is the unstrained surface energy, $\sigma_{ij}$ the surface
stress tensor, $\epsilon_{ij}$ the strain tensor, and $S_{ijkl}$ the tensor of
second order stresses. The calculation of the first order surface stress is
done as follows.  We calculate the surface energy of a slab for various
lattice constants in the range of $\pm 4$ \%. The strained surfaces do not
have to be relaxed again after straining because the relaxation energy is of
second order in the strain. The energies of the strained surfaces are fitted
to a polynomial from which we extract the linear coefficient of surface energy
as a function of the strain. In all calculations we find that the components
of the surface stress tensor are tensile. Compared to the Si (100) surface its
value has the same order of magnitude. \cite{garcia:93} For the (110) surface
$\sigma_x$ and $\sigma_y$ denote components parallel to [001] and [1\=10]
respectively. For the (111) and (\=1\=1\=1) surface the surface stress tensors
are isotropic due to the three fold symmetry of the surface.  For the (110)
surface we also evaluated the strain derivative of the surface stress by
compressing the surface isotropically up to 12 \%. The surface energy of the
strained surface is computed by subtracting the corresponding energy of
strained bulk from the total energy of the relaxed strained slab. The surface
energy as a function of the strain is fitted to a second order polynomial. The
second order coefficient gives the sum of the derivatives of the surface
stress $S_{1111} + 2 S_{1122} + S_{2222}$ which is equal to $-0.5$
eV/{\AA}$^2$. Straining the InAs (110) surface epitaxially to the GaAs
lattice constant would result in a contribution from the second order terms to
the surface energy smaller than 2 meV/{\AA}$^2$, while the linear term amounts
to $\sim 6 $ meV/{\AA}$^2$. Thus we neglect in the following the second and
higher order corrections to the surface energy.
 
\begin{figure}[tb]
  \begin{center}
    \epsfig{figure=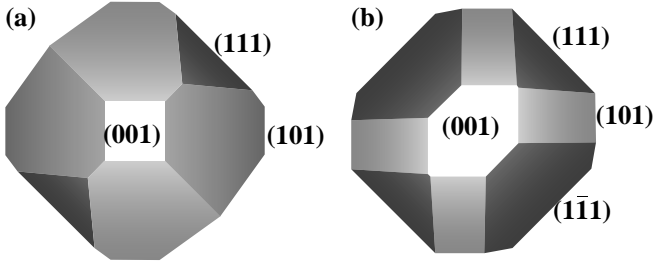, width=\linewidth}
  \end{center}
  \caption{The equilibrium crystal shape of InAs in (a) In-rich and (b) 
    As-rich environment. The Miller indices of the surfaces are noted.}
  \label{ecs}
\end{figure}

Under thermal equilibrium conditions, the shape of a large InAs crystallite is
given by the condition of lowest free energy. This can be obtained for zero
temperature by applying the Wulff construction using the surface energies of
Tab. \ref{tab}. The resulting shape is shown in Fig.\ \ref{ecs}.
As this figure is constructed using only the surface energies of the \{110\},
\{100\}, \{111\}, and \{\=1\=1\=1\} orientations, we note that in general also
higher index surfaces might be present, but the low Miller-index surfaces are
expected to remain clearly prominent. Fig.\ \ref{ecs} shows that under As-rich
conditions all four considered surface orientations coexist on the equilibrium
crystal shape.  This result is in agreement with the shape of large, and thus
presumably fully relaxed, InAs islands grown on a GaAs substrate by Steimetz
{\em et al}.\ \cite{steimetz:97} using metal-organic vapor-phase epitaxy.
Under In-rich condition the \{\=1\=1\=1\} facets do not exist because they are
energetically unfavorable. This is probably due to the fact that we did not
consider the $(\sqrt{19} \times \sqrt{19})$ structure.

\section{Equilibrium Shape of Isolated Quantum Dots}

The equilibrium shape of a strained coherent island of a given number of atoms
is determined by the minimum of its total energy with respect to its shape.
To determine the optimum island shape as a function of volume we follow the
procedure already outlined in Ref.\ \onlinecite{pehlke:97}. As described in
Section I the total energy is evaluated by summing the elastic energy, the
strain renormalized surface energy and the edge energy. 

Accurate values of the edge energies are not known. We examined the influence
of edge energies by calculating the equilibrium island shapes assuming the same
value of the edge energy for all types of edges. We found that the island
shape was not influenced as long as edge energies are smaller than 100
meV/{\AA} for a quantum dot of 10,000 atoms, mainly because the edge energies
only scale with $V^{1/3}$. Heller {\em et al}.\ \cite{heller:93} measured the
energies of steps on the GaAs (100) surface to be 4 meV/{\AA} and 13 meV/{\AA}
for the two different types of steps. Recently, Kratzer and Scheffler
\cite{kratzer:97} computed a value of 25 meV/{\AA} for the one type. Edge
energies should be of comparable size or even smaller and thus do not play a
role for the island shape. Therefore, we neglect the edge energies in
the following analysis.

\begin{figure}[tb]
  \begin{center}
    \epsfig{figure=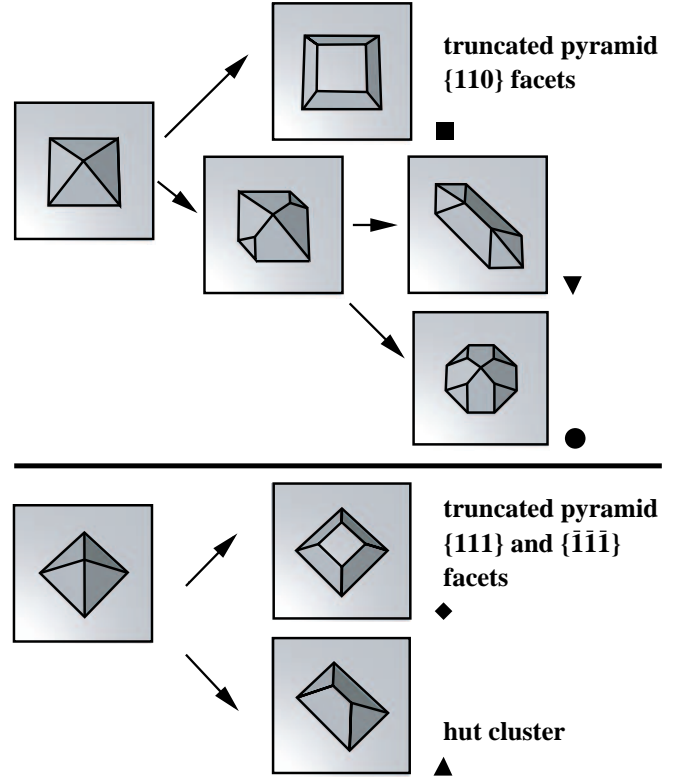, width=\linewidth}
  \end{center}
  \caption{The different island shapes which are investigated. They consist
    out of \{101\}, \{001\}, \{111\}, and \{\=1\=1\=1\} facets.}
  \label{overview}
\end{figure}

\begin{figure}[tb] 
  \epsfxsize=\linewidth 
  \epsfbox{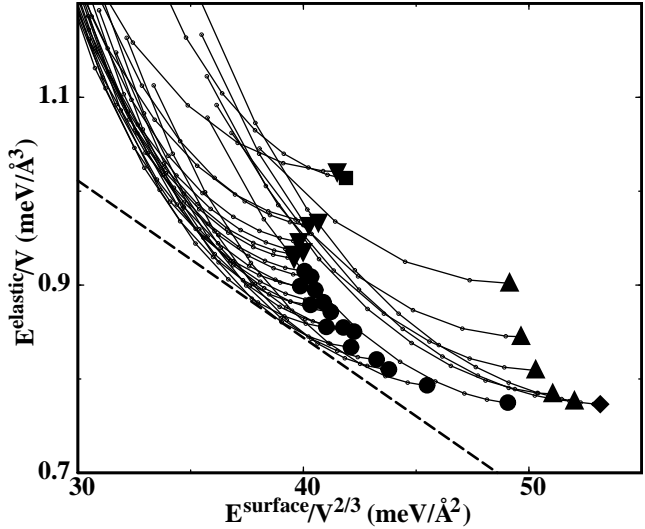}
  \caption{The elastic energy per volume $E^{\rm elastic}/V$ versus the
    surface energy per area $E^{\rm surface}/V^{2/3}$ for InAs islands.  The
    symbols refer to the shapes displayed in Fig.\ \protect\ref{overview}:
    {\em Square:} square based pyramid with four \{101\} facets.  {\em
    Diamond:} square based pyramid with two \{111\} and two \{\=1\=1\=1\}
    facets.  {\em Triangles up:} huts with two \{111\} and two \{\=1\=1\=1\}
    facets.  {\em Triangles down:} square based \{101\} pyramids with
    \{\=1\=1\=1\} truncated edges.  {\em Dots:} islands with four \{101\}, two
    \{111\}, and two \{\=1\=1\=1\} facets. The open circles denote the
    corresponding truncated islands which are connected by the full lines. The
    dashed line is the curve of constant total energy $E_{\rm elastic} +
    E_{\rm surface}$ that selects the equilibrium shape for the volume $V =
    2.14 \times 10^5 {\rm {\AA}}^3$.}
  \label{elastic}
\end{figure}

Due to the different scaling properties of the elastic and surface energy, the
optimum island shape depends on the volume.  We consider all possible island
shapes which have low-index surface facets. An overview is shown in Fig.\
\ref{overview}.
If surfaces with other orientations and a smaller slope than the \{101\}
facets would appear on the islands, these surfaces would facet into the
thermodynamically stable orientations \{101\}, \{001\}, \{111\}, and
\{\=1\=1\=1\}. We calculate the total energies of islands bounded by \{101\},
\{111\}, and \{\=1\=1\=1\} facets. The filled symbols in Fig.\ \ref{elastic}
are obtained using results from full finite element calculations, whereas the
small dots denote the truncated mesa shaped islands where the elastic energies
are derived from the simple analytical approximation.
The results of the elastic calculations are combined with the {\em ab initio}
surface energies. Because the side facets of the islands are strained we
include the first order correction of the surface energy due to the
strain. For this we use the {\em ab initio} stress tensors from Tab.\
\ref{tab} and combine it with the strain field at the surface from the finite
element calculation.  We have used the surface energy of the unstrained InAs
(100) surface for the almost fully relaxed (100) top facets of the quantum
dots, and the surface energy of an isotropically strained wetting layer for
the area covered by the quantum dots. The contribution of the elastic and
strain renormalized surface energies are displayed in Fig.\ \ref{elastic},
both of them divided by the their respective scaling factors.

\begin{figure}[tb]
  \epsfxsize=\linewidth
  \epsfbox{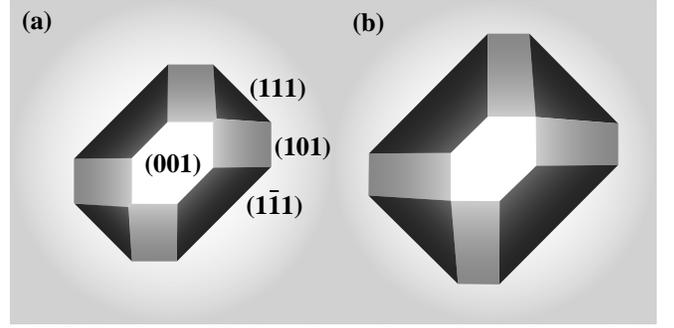}
   \caption{The equilibrium shape of a strained coherent InAs islands in
     As rich environment at two different volumes, (a) $V \approx 
     2 \times 10^5 {\rm \AA}^3$ (10,000 atoms), (b) $V \approx 4 \times
     10^5 {\rm \AA}^3$ (20,000 atoms).}
  \label{islands}
\end{figure}

In Fig.\ \ref{elastic} the optimum island shape is determined by the point
where the line of constant total energy touches the manifold of the island
energies from below. Therefore, the equilibrium island shapes for
all volumes are given by the lower envelope of the manifold of the island
energies. In Fig.\ \ref{islands} the equilibrium island shapes determined by
this method are shown for two different volumes.
The elastic energy scales with the volume $V$ whereas the strain renormalized
surface energy increases with the volume like $V^{2/3}$, because the strain
field of the side facets is invariant to the scaling. Due to their scaling the
surface energy dominates at small volume and the elastic energy gains more
importance at large volume.  Thus, larger islands are steeper than smaller
ones. As the \{111\} and \{\=1\=1\=1\} facets are steeper than the \{101\}
facets they become more prominent on larger islands. Therefore, the island
shape is not fixed, but changes continuously with the volume. This change in
the shapes implies that the simple scaling laws which are valid for a fixed
shape do not apply.

The influence of the surface stress is in such a way that surface energies of
the \{111\} and \{\=1\=1\=1\} facets are further lowered as compared to the
\{101\} facets. Roughly 30 \% of the surface of each facets are strained while
the rest is completely relaxed. This means that the \{101\} facets are lowered
due to the strain by $\Delta \gamma = -1.7$ meV/{\AA}$^2$ whereas the \{111\}
and \{\=1\=1\=1\} facets are lowered by $\Delta \gamma = -2.1$ meV/{\AA}$^2$
and $\Delta \gamma = -4.1$ meV/{\AA}$^2$, respectively, i.e., the
\{\=1\=1\=1\} facets are lowered 6 \% more in surface energy than the \{101\}
facets.  Thus, the \{\=1\=1\=1\} faces dominate even more due to the surface
strain. Although influenced by strain effects the equilibrium island shape
remains similar to the ECS.  

Recent atomic force microscopy (AFM) studies by Georgsson {\em et al.}\
\cite{georgsson:95} of uncapped (not overgrown) InP islands on a GaInP
substrate corroborate our theoretical predictions. The experimental island
shape displays the same facets as our calculated island shape. Although InP
differs from InAs, {\em ab initio} results \cite{moll:97} yield a similar
behavior for the surface energies of InP as for InAs. Therefore, we expect to
obtain a similar theoretical shape for InP islands. For InAs, however, the
experimentally observed shapes differ from ours and show various different
shapes. Moison {\em et al.}\ \cite{moison:94} observe \{410\} to \{110\}
facets and Ruvimov {\em et al.}\ \cite{ruvimov:95} only \{110\}
facets. Whereas Leonard {\em et al.}\ \cite{leonard:94} report their island
shapes as planoconvex lenses with a radius to height aspect ratio of about
two. The great diversity of experimental island shapes and the difference to
ours indicates that the respective growth conditions do not represent
thermodynamic equilibrium but are driven by kinetics.  Thus, kinetic effects
such as Jesson {\em et al.}\ \cite{jesson:96} suggested may play a role. These
kinetic effects could not only affect the distribution of the island size but
also the shape of the islands.  However, it should be possible to achieve
thermodynamic equilibrium by choosing appropriate experimental conditions.
Alloying of the quantum dots with the GaAs matrix will also play a role and
might affect the shape of the quantum dots. \cite{rosenauer:97}

To gain further insight into the shape of the islands, more experimental
investigations should be performed such as high resolution STM or AFM of the
side facets. This could help the theory for further investigations.

\section{Acknowledgments}

This work was supported in part by the Sfb 296 of the Deutsche
Forschungsgemeinschaft.

\end{document}